\documentclass[%
 reprint,
 amsmath,amssymb,
 aps, prl,
]{revtex4-2}

\usepackage{graphicx}
\usepackage{bm}
\usepackage{dutchcal}
\usepackage{tensor}
\usepackage{xcolor}

\newcommand{\dderiv}{\mathrm{d}}
\newcommand{\Rtensor}{R^{0}_{\hphantom{0}i0j}}

\newcommand*{\unit}[1]{\,\mathrm{#1}}

\DeclareMathAlphabet{\mathpzc}{OT1}{pzc}{m}{it}

\begin{document}


\title{
Gravitational Glint: detectable gravitational wave tails from stars and compact objects
}

\author{Craig Copi$^1$}
 \email{cjc5@case.edu}
 \author{Glenn D. Starkman$^{1,2}$}
  \email{gds6@case.edu}
\affiliation{%
$^1$Department of Physics/CERCA/ISO,
Case Western Reserve University, Cleveland OH 44106, USA
}%
\affiliation{%
$^2$Department of Physics, Imperial College London,  SW7 2AZ, UK
}%

\date{\today}

\begin{abstract}
Observations of a merging neutron star binary in both gravitational waves, by the Laser Interferometer Gravitational-wave Observatory (LIGO),
and across the spectrum of electromagnetic radiation, by myriad telescopes,
have been used to show that gravitational waves travel in vacuum at a speed that is indistinguishable from that of light to within one part in a quadrillion.  
However, it has long been expected mathematically that, when electromagnetic or gravitational waves travel through vacuum in a curved spacetime, the waves develop “tails” that travel more slowly.  
The associated signal has  been thought to be undetectably weak.  
Here we demonstrate that gravitational waves are efficiently scattered by the curvature sourced by ordinary compact objects---stars, white dwarfs, neutron stars, and planets---and certain candidates for dark matter, populating the interior of the null cone.
The resulting “gravitational glint”
should imminently be detectable, and be recognizable (for all but planets) as briefly delayed echoes of the primary signal emanating from extremely near the direction of the primary source. 
This opens the prospect for using GRAvitational Detection And Ranging (GRADAR) to map the Universe and conduct a comprehensive census of massive compact objects, and ultimately to explore their interiors.
\end{abstract}

\maketitle

A key discovery \cite{Michelson:1887zz} that underpinned the development of modern physics  \cite{Einstein:1905ve} is that electromagnetic (EM) radiation travels in vacuum at a unique speed $c$ no matter the velocity of any inertial observer. 
Recently, it was demonstrated, through simultaneous observation of gravitational waves (GWs) and EM emissions from a merging neutron-star binary \cite{LIGOScientific:2017vwq}, that GWs in vacuum travel at that same speed.
Nevertheless, it has long been understood \cite{hadamard1923lectures}, though not widely known,
that  General  Relativity (GR) generically predicts that massless scalar, EM, and GWs propagate both on {\emph{and inside}} the null cone---the Green's function (GF) of the relevant wave operator has support not just on the null cone but also on its interior.
Only in homogeneous, conformally flat, even-dimensional spacetimes---such as $3+1$d Minkowski spacetime and flat Lemaitre-Friedmann-Robertson-Walker spacetime---does the GF vanish inside the null cone and massless fields travel exclusively at $c$.
Propagation on the interior of the null cone is known as the ``tail effect.''  
To our knowledge, this tail signal has never been directly, nor indirectly, detected.

This letter identifies exciting observational prospects for the tail signals of GWs like those  observed \cite{LIGOScientific:2016aoc} emanating from merging black-hole and neutron-star binaries.  These propagate in a space  populated by massive ``perturbers'' of the spacetime geometry, such as stars and their post-fusion remnants---white dwarfs and neutron stars. 
Primarily because observed GWs are much longer wavelength than observable EM waves, we find that the GW tails are imminently observable, more so  than their EM counterparts predicted in \cite{Copi:2020qur}.

While the early-time and late-time behaviours of the tail in Minksowski space perturbed by a mass have been studied previously, here we present the new piece that arises from probing the internal structure of a  perturber.  
As in the EM case \cite{Copi:2020qur}, this ``middle-time tail'' (MTT) proves significantly stronger than the early and late pieces for perturbers such as stars and stellar remnants, situated outside but near the Einstein radius along the line-of-sight (LOS) to any GW source.
MTT signals may be detectable in existing GW detectors, possibly in archival data.
They could potentially reveal the distribution, and even structure, of such perturbers. 

The nature of the tail depends on the geometry, and thus on the matter sourcing it.
A calculationally tractable, observationally relevant case is an isolated, spherical, compact  object---such as a star or its non-black-hole compact remnant---near the LOS between a source and an observer in otherwise-empty (Minkowski) spacetime.
Recently, we examined this situation for the propagation of EM waves \cite{Copi:2020qur}, using the perturbative-GF-approach developed in \cite{Pfenning_2002,Chu:2011aa}, and the specific calculation of the GF of \cite{Chu:2020aa}.
Given this perturbative approach, we considered only ``weak-(gravitational)-field'' perturbers---i.e., not black holes, and neutron stars less-reliably.
Because EM waves with wavelength $\lambda \gtrsim 100\unit{m}$ do not penetrate the interplanetary medium, we considered $\lambda$ much smaller than all other scales in the problem.

\begin{figure}
    \centering
    \includegraphics{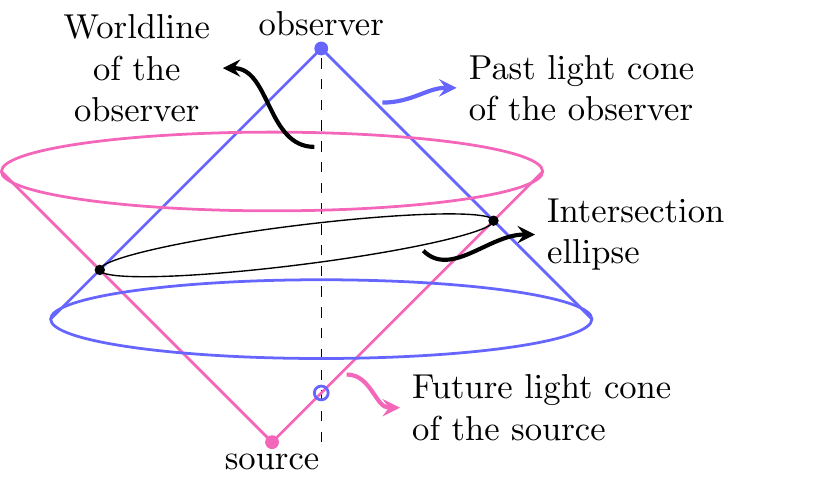}
    \caption{Spacetime diagram of a source emission and observation.
    In the usual description of their connection---the 0th  order contribution to the GF---the source emits a signal that the observer receives when their worldline crosses the forward null cone of the source (hollow circle). The next-order contribution represents the lightcone signal interacting with the perturbed spacetime geometry at a later  time. Such ``scattering'' events lie on the past null cone of the observer at some even later observation time. The observer adds up the contributions to the ``scattered'' signal from the ellipse (ellipsoid in 3+1-dimensions) of points on the intersection of their past null cone and the forward null cone of the source.}
    \label{fig:lightcones}
\end{figure}

Expanding the  GF in a perturbation series around a (conformally) flat-spacetime GF, the leading-order tail contribution can be  conceptualized as represented in Fig.~\ref{fig:lightcones}:
a (brief) EM wave or GW  signal travels out from a source along the null cone;  it later interacts with the perturbed geometry; it then propagates from the ``scattering site'' to the observer, again along a null path, arriving after the direct  null cone signal.
The observer detects the superposition of all such signals arriving simultaneously, which therefore have interacted with the geometry on an  ellipsoid  with foci at the source and observer.

The principal new insight of \cite{Copi:2020qur} was that the EM tail signal is dominated by the epoch during which the ellipsoid of equal-time scattering sites intersects the interior of the perturber.
Previous treatments had taken the perturbing matter source to be a delta-function.
Given their  weak-field approximation, this precluded treating the effects of the region inside a physical perturber, which requires resolving the delta function.
The density profile of the perturber $\rho(r)$ proves crucial to assessing the detectability of the tail signal.
For illustrative purposes, we study a spherically symmetric perturber with 
\begin{equation} 
    \label{eqn:rhoofr}
    \rho(r) = 
    \left\{
      \begin{array}{cc}
        \rho_\mathrm{central}\left( 1 - \frac{r^2}{a^2}\right)^p ,& r\leq a\\
        0 ,& r > a .
      \end{array}
    \right.
\end{equation}
$a\gg r_S=2G_N M_P/c^2$, the Schwarzschild radius, enforces the weak-field approximation.
For our GF method 
to yield physically meaningful results,  $\rho(r)$ must be sufficiently smooth.
We found that $p\geq2$ is required for EM waves, but $p\geq4$ is needed for the GW calculations presented here.

\begin{figure*}[!ht]
    \centering
    \includegraphics{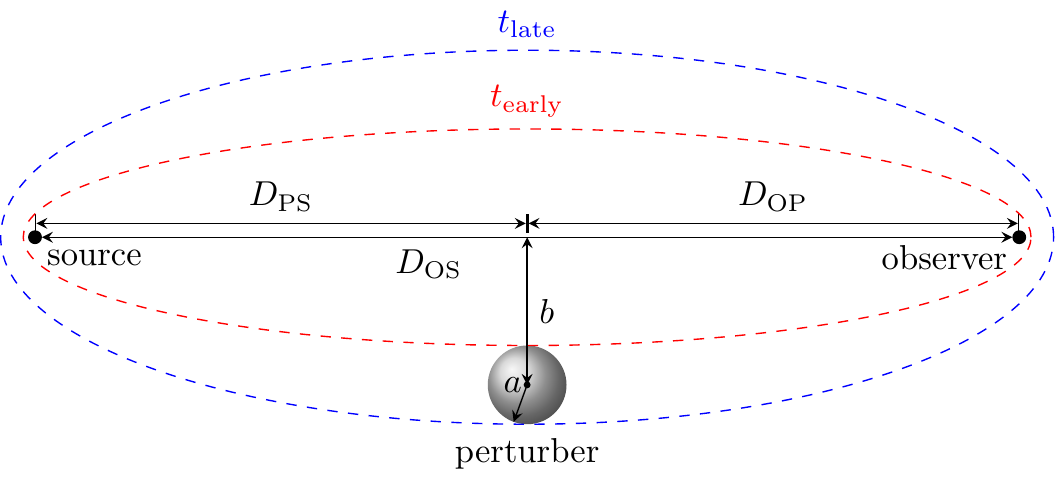}
    \caption{A perturber  radius $a$ a distance $b\gg a$ from the midpoint of the LOS ($D_{\mathrm{PS}}=D_{\mathrm{OP}}=D_{\mathrm{OS}}/2\gg b$) between a GW source and an observer.
    The surfaces of constant total source-point-observer distance (in the flat, static, background geometry) are nested ellipsoids, which project to ellipses in the source-observer-perturber plane.
    All points on each  ellipsoid  contribute to the signal at a fixed time, at leading order in the weak-field expansion of the GF.
    The $t_{\mathrm{early}}$ ($t_{\mathrm{late}} $) ellipsoid osculates the surface of the perturber closest to (furthest from) the LOS.
    Before $t_{\mathrm{early}}$ the perturber is outside  the ellipsoid, and the observer receives the early-time tail, which vanishes at leading order. After $t_{\mathrm{late}}$ they receive the small late-time tail.  Between $t_{\mathrm{early}}$ and $t_{\mathrm{late}}$ the ellipsoid overlaps the perturber, and the observer receives the much stronger MTT.
    Drawing not to scale.}
    \label{fig:coordinates}
\end{figure*}

For EM, we confirmed some well-known results \cite{Copi:2020qur}. 
First, there is no ``early'' tail signal, i.e., no light would be observed after the null cone signal and before $t_{\mathrm{early}}$---the first time a signal could propagate at $c$ from the source to anywhere on the physical surface of the perturber and thence to the observer. 
Second,  there is \cite{1995PhRvL..74.2414C} a ``late-time tail'' after $t_{\mathrm{late}}$---the last time that a signal could make such a journey (see Fig.~\ref{fig:coordinates}).
The late-time tail is  extremely small at $t_{\mathrm{late}}$, and falls rapidly thereafter.

Previous treatments ignored the MTT between $t_{\mathrm{early}}$ and $t_{\mathrm{late}}$. 
By resolving $\rho$, and thence $g_{\mu\nu}$, inside the perturber with (\ref{eqn:rhoofr}), we were able to compute the EM MTT;  
while small, it was vastly larger than the early or late-time tails.
Unfortunately, the EM MTT is only accessible for transparent perturbers, eliminating dense baryonic objects like stars, but potentially including dark-matter overdensities, such as axion minihalos, and more diffuse baryonic perturbers, such as globular clusters. However, our calculation would not strictly apply.

An important lesson of the EM calculation was that, to access the larger MTT signal, one favors transparent perturbers and long-wavelength sources.
These are both natural for GWs---only black holes absorb substantial fractions of an incident GW flux. 
Meanwhile LIGO sensitivity peaks around $100\unit{Hz}$, corresponding to $\lambda_{\mathrm{GW}}\simeq3000\unit{km}$. 
Pulsar timing arrays and the planned LISA observatory are sensitive to even longer wavelengths.

Here we calculate the amplitude and character of that GW MTT signal and compare it to its corresponding null cone signal.
We show that compact concentrations of matter near the LOS produce nearly faithful echoes of primary signals.
We argue that detectable null cone GW signals will be accompanied by detectable MTTs sufficiently frequently to make them exciting new probes of the contents of the Universe and a new test of GR.

A GW is characterized by a small deviation, $h_{\mu\nu}\equiv g_{\mu\nu}-\bar{g}_{\mu\nu}$, of the exact metric $g_{\mu\nu}$ from the background geometry, $\bar{g}_{\mu\nu}$, through which it propagates.
$h_{\mu\nu}$ encodes the amount by which spacetime is altered by the propagating wave. 
We derive the inside-the-null-cone piece, $h^{(\mathrm{tail})}_{\mu\nu}$, observable as a result of a distant GW source,
and a static, spherically symmetric, compact, weak-field mass distribution.
We use the same perturbative approach as \cite{Copi:2020qur}.
To first order in the perturbation, the gravitational field propagating from $x$ to $x'$ is \cite{Chu:2011aa}
\begin{widetext}
\begin{align}
h_{\mu \nu}(x)&= \int \dderiv^{4} x' \sqrt{\vert g\vert} G_{\mu \nu \alpha' \beta'} J^{\alpha' \beta'} + 
    \int \dderiv^{3} x' \sqrt{\vert g\vert} 
        \left[
            G_{\mu \nu \alpha' \beta'} \tensor{P}{^{\alpha'}^{\beta'}_{\rho'}_{\epsilon'}} \nabla^{0'} h^{\rho'\epsilon'}
            -
            \left(\nabla^{0'}G_{\mu \nu \alpha' \beta'}\right) \tensor{P}{^{\alpha'}^{\beta'}_{\rho'}_{\epsilon'}} h^{\rho'\epsilon'}
        \right]\,, 
\end{align}
\end{widetext}
where 
$G_{\mu \nu \alpha' \beta'}$ is the GF of  $h_{\mu\nu}$ (cf. \cite{Chu:2011aa, Chu:2020aa} for details) and
\begin{equation}
   \tensor{P}{^{\alpha'}^{\beta'}_{\rho'}_{\epsilon'}} 
    \equiv \frac{1}{2}
    \left(
     \delta_{\rho'}^{\alpha'}\delta_{\epsilon'}^{\beta'} 
     +
     \delta_{\epsilon'}^{\alpha'}\delta_{\rho'}^{\beta'} 
     -
     g^{\alpha' \beta'} g_{\rho'\epsilon'} 
    \right).
\end{equation}

In de Donder gauge, 
$G_{\mu\nu\alpha\beta}(x,x')$ for a compact, spherical perturber can be written in terms of four derivatives of the same scalar two-point function $A(x,x')$ that is the basis of the EM GF \cite{Copi:2020qur}.
$G_{\mu\nu\alpha\beta}$ can be used to compute $h_{\mu\nu}$ at the observer in at least two distinct ways. In the ``source approach,''  the GF is integrated against
\begin{equation}
    J^{\alpha' \beta'} \equiv 16\pi \tensor{P}{^{\alpha'}^{\beta'}_{\rho'}_{\epsilon'}}  T^{\rho'\epsilon'}
\end{equation}
of the GW source.  
In the  ``initial-value-problem (IVP) approach,'' the  GW perturbation generated by the source $h^{\mathrm{(IVP)}}_{\mu\nu}$, and its gradient, are evaluated well after the event that generated the GW but long before they have propagated far on the null cone. 
$h^{\mathrm{(IVP)}}_{\mu\nu}$ and its gradient are then integrated against the GF and its gradient \cite{Chu:2011aa}.
To its advantage, the IVP approach can take as input any compact-support GW, including the full numerical templates developed with much effort for the coalescing binary black holes and neutron stars being detected by LIGO and the Virgo observatory.
We compared the results of these two  approaches to validate our algebraic and numerical calculations. 

We take as a representative gravitational source a binary, consisting of two equal masses $m_{\mathrm{b}}$  in a circular orbit of radius $R_{\mathrm{b}}$ about their mutual center-of-mass, with angular frequency  $\Omega_{\mathrm{b}} = \sqrt{2G_Nm_{\mathrm{b}}/R_{\mathrm{b}}^3}$. 
This binary emit GWs with frequency $\nu_{\mathrm{GW}}=\Omega_{\mathrm{b}}/\pi$.
We treat this system in the Newtonian limit (weak field and slow orbital speed).
Although the black-hole and neutron-star binaries observed by GW observatories are  strong-field and  high-velocity, this only affects the signal emitted at the source, not the changes on the GW propagation due to the perturber.  
Therefore,  we expect our conclusions to apply equally to the signals from those binaries.

$h_{\mu\nu}$ is not itself an observable. 
GW observatories are actually sensitive to  $\ddot{h}^{\mathrm{TT}}_{ij}$,  over-dots being time derivatives, and $i$ and $j$ referring to spatial components.
TT stands for transverse and traceless, meaning $h_{ij}$ has been cast into a form that makes manifest the interpretion of GWs as oscillations in the spatial geometry perpendicular to the direction of propagation, with only two polarizations, in direct analogy to EM radiation.
This itself can be written \cite{Maggiore:2007} in terms of specific components of the Riemann tensor, $\ddot{h}^{\mathrm{TT}}_{ij}=-2 c^2 \Rtensor$.

\noindent\paragraph*{Simplifying assumptions}

Calculation of $\Rtensor$ from $h_{\mu\nu}$ adds two more derivatives to the computation of the observable tail signal. 
The proliferation of terms from six derivatives of $A$ greatly complicates the calculation of $\Rtensor$, so agreement of the two GF approaches is reassuring.  Even so, the calculation would remain uninstructive if not for several simplifying observations and assumptions that enable us to present simple, numerically justified, analytic expressions for the tail signal in useful regimes.
These assumptions pertain to the geometry of the source-perturber-observer system as shown in Fig.~\ref{fig:coordinates}:\newline
    \noindent$\bullet$ The Schwarzschild radius of the perturber is much less than its size, $r_S \ll a$; though for a neutron star $a/r_S\simeq3$.   
\newline
    \noindent$\bullet$ The size of the perturber is much less than its perpendicular distance to the LOS:  $a\ll b$.\newline
    \noindent$\bullet$ $b$ is much less than half the length of the LOS:  $b\ll \ell  $.\newline
    \noindent$\bullet$ GW are naturally in a long-wavelength regime, 
    \begin{equation}
        \lambda_{\mathrm{GW}}/c =\nu_{\mathrm{GW}}^{-1}\gg \tau_{\mathrm{middle}}\equiv (t_{\mathrm{late}}-t_{\mathrm{early}}).
    \end{equation}
    $\tau_{\mathrm{middle}}$ is the width of the middle-time portion of the GF.  
    $\tau_{\mathrm{middle}}\ll \nu_{\mathrm{GW}}^{-1}$ reduces the distortion of the null cone signal, so that the MTT is nearly its echo.\newline
    \noindent$\bullet$ It would be useful for $\Delta t_{\mathrm{early}}\equiv t_{\mathrm{early}}-t_{\mathrm{null}}\gtrsim \nu_{\mathrm{GW}}^{-1}$, so that the MTT comes noticeably after  the null cone signal,  but this is not a requirement.\newline
    \noindent$\bullet$ To further simplify the calculation, we take the observer-perturber distance along the LOS $D_{\mathrm{OP}}$  to be half the observer-source distance  $D_{\mathrm{OS}}$, so $\ell   \equiv D_{\mathrm{OP}} = D_{\mathrm{PS}} = D_{\mathrm{OS}}/2$. 
    (Although we consider a Minkowski background, if corrections due to cosmic expansion are relevant,
    these should be interpreted as angular diameter distances.)
    Though we have not carefully tested the limits of validity of our results with respect to deviations from this special case, they appear to hold to leading order in $\Delta \ell  /\ell  $, so approximately over most of the  LOS. 

To leading order in the resulting small quantities, $c\Delta t_{\mathrm{early}}\simeq b^2/\ell  $ and $c\tau_{\mathrm{middle}}\simeq 4 a  b/\ell$. 
In summary, we use the following hierarchies of scale to implement our algebraic and numerical evaluation of the signal:
\begin{equation}
    \label{eq:smallquantities}
    \frac{4 a  b}{\ell  } \ll \frac{c}{\nu_{\mathrm{GW}}} \lesssim \frac{b^2}{\ell};
    \quad r_S \ll a \ll b \ll \ell .
\end{equation}

Analytically to leading order in the various small quantities, for the weak-field circular equal-mass binary 
\begin{equation}
    \label{eqn:MTTSignal}
    {\vert\ddot{h}^{\mathrm{TT;middle}}_{ij}\vert} = \frac{32 G_N^2 m_{\mathrm{b}} M_P f_P  R_{\mathrm{b}}^2 \Omega_{\mathrm{b}}^4}{b^2 c^2} ,
\end{equation}
so long as $b/x_E\equiv n_E\gtrsim\mathrm{few}$.
Here
\begin{equation}
    x_E \equiv \sqrt{
    \frac{4 G_N  M_{P}}{c^2} \frac{D_{\mathrm{OP}}D_{\mathrm{PS}}}{D_{\mathrm{OS}}}} = \sqrt{\frac{2 G_N M_P \ell}{c^2}  }\,
\end{equation}
is the Einstein radius, inside which a perturber causes strong lensing, i.e., multiple images of a source.
For $n_E\lesssim \mathrm{few}$, our calculation becomes unreliable.
$f_P$ is a numerical factor depending on the density profile of the perturber, with $f_P=1$ for $p=4$ in  \eqref{eqn:rhoofr}.

It is useful to compare (\ref{eqn:MTTSignal}) to the  amplitude of the null signal from such a binary (e.g., \cite{Carroll:2003})
\begin{equation}
    \label{eqn:LCSignal}
    {\vert\ddot{h}^{\mathrm{TT;null}}_{ij}\vert} =  \frac{32 G_N  m_{\mathrm{b}} R_{\mathrm{b}}^2 \Omega_{\mathrm{b}}^4}{\ell c^2 }.
\end{equation}
These amplitudes are source-dependent.
More crucially, their ratio reduces to
\begin{equation}
    \label{eqn:StoN}
    \mathcal{R}^{\mathrm{\mathrm{mid/null}}} \equiv
    \frac{\vert\ddot{h}^{\mathrm{TT;middle}}_{ij}\vert}{\vert\ddot{h}^{\mathrm{TT;null}}_{ij}\vert} =  \frac{G_N M_{P} f_P \ell  }{b^2 c^2} =\frac{f_P}{n_E^2}.
\end{equation}

Equation (\ref{eqn:StoN}) is a most exciting result.
It  implies that GW sources that have been observed at high signal-to-noise can cause detectable glint from perturbers within some modest number of Einstein radii from the LOS\@.
The expectation that such reflections would be weak because gravity is weak is realized in the factor of $G_N$ in the numerator, or alternately in the very small ratio of the perturber's Schwarzschild radius to its distance from  the LOS, $2G_N M_P/b~c^2$.  
However this is compensated for by the length of the LOS compared to that distance, $\ell  /b$.

\noindent\paragraph*{Glint direction}
Ratios of  components of $\ddot{h}^{\mathrm{TT}}_{ij}$,
determine the direction from which we interpret the GW arriving.
As for EM waves, to leading order the MTT  emanates from the perturber,  probably undetectably displaced from the source  with any anticipated detectors.

\noindent\paragraph*{Glint timing}
Because $c\tau_{\mathrm{middle}}\ll\lambda_{\mathrm{GW}}$, the middle-time segment of the GF is  a narrow kernel against which the null cone signal from the source is convolved.
As (\ref{eqn:StoN}) is $\lambda_{\mathrm{GW}}$-independent, the leading-order MTT is a faithful echo of that signal, delayed by  $\Delta t_{\mathrm{early}}$.

The close temporal and directional association of the glint  with the brighter null cone signal should facilitate glint detection.  
Since the leading-order MTT is a faithful echo of the null-cone signal,  arriving shortly thereafter from nearly the same direction, glints  need not be detected on a stand-alone basis out of the noise; but should rather be sought in association with the primary signal.
 
A glint well-separated from the primary signal would be easiest to detect; it would have $\Delta t_{\mathrm{early}}$ larger than the GW period.
Expressed in terms of the GW frequency, this means $\nu_{\mathrm{GW}} \Delta t_{\mathrm{early}} \gtrsim1$, where
\begin{equation}
    \label{eqn:nu_deltatearly}
    \nu_{\mathrm{GW}} \Delta t_{\mathrm{early}} \simeq 10^{-3} n_E^2 \frac{M_P}{\mathrm{M_\odot}} \frac{\nu_{\mathrm{GW}}}{100\unit{Hz}}\,.
\end{equation}
Because a continuous single-frequency signal plus a faithful  echo is indistinguishable from a time-delayed signal of higher amplitude,
the detectability of each glint depends on the precise frequency and temporal structure of the primary signal. 
Glints of a   well-characterized non-monochromatic primary signal may well be detectable with  $\nu_{\mathrm{GW}} \Delta t_{\mathrm{early}}\ll1$.
The most powerful GW signals, coalescing black-hole binaries, have well-characterized, rapidly evolving frequency spectra during the chirp and ringdown phases. 

As a function of the glint-to-null-cone signal ratio
\begin{equation}
    \label{eq:tearlyvssignal}
    \nu_{\mathrm{GW}}\Delta t_{\mathrm{early}} \simeq \frac{10^{-3}}{\mathcal{R}^{\mathrm{mid/null}}}
    f_P 
    \frac{M_{\mathrm{stellar}}}{\mathrm{M_\odot}} \frac{\nu_{\mathrm{GW}}}{100\unit{Hz}}\,.
\end{equation}
Since weaker glints have longer time delays (but still short enough to be closely associated with the null-cone signal), this should again aid in their detection.
Nevertheless , Eq.~(\ref{eq:tearlyvssignal}) suggests that it will be difficult to use the leading-order glint signal to discover and characterize objects that are  much less than a solar mass. This includes  abundant known objects like planets, which 
may best be identified as perturbations on stellar glints.

\noindent\paragraph*{Glint Statistics}
The expected number of uniformly distributed perturbers of mass $M_P$ lying
within $n_E$ Einstein radii of a LOS, but outside $x_E$, is (in the approximation that we evaluate $x_E$ at the middle of the LOS):
\begin{equation}
    \label{eq:NofnE}
    N(n_E) = \frac{3}{4}  \Omega_P  (H_0 \ell  )^2 \frac{\Delta\ell}{\ell  } (n_E^2-1) .
\end{equation}
Here $\Omega_P$ is the fraction of the cosmological energy density  in such perturbers, and $\Delta\ell/\ell  $ is the fraction of the LOS over which  (\ref{eqn:MTTSignal}) holds.  
We expect $\Delta\ell/\ell  \simeq1$, but detailed characterization of the variation of the tail signal along the LOS is reserved for future work.

With $\Omega_{\mathrm{stars}}\gtrsim 0.003$, and 
typical coalescing-black-hole-binary GW sources at $\simeq10\%$ of the Hubble distance
\begin{equation}
    \label{eqn:Nstellar}
N_{\mathrm{star}}(n_E )\gtrsim 1.5 \times 10^{-3} n_E^2 = \frac{1.5 \times 10^{-3}}{\mathcal{R}^{\mathrm{mid/null}}}.
\end{equation}
We therefore expect one glint $1/3$ as bright as the source for every $225$ sources  observed.  This is approximately once every three years at the current LIGO event rate, so one or more such glints may be found in archival data.
Expected improvements to LIGO, for example squeezed light sources \cite{LIGORDLI27:online}, may increase this event rate ten-fold.

Eq.~(\ref{eq:NofnE}) assumes a homogeneous distribution of perturbers, whereas stars are concentrated in galaxies.  This may have a significant impact on glint statistics, especially the prospects for multiple glints from the same primary source when the LOS passes through a galaxy.

\noindent\paragraph*{No confusion with strong lensing}
Strong lensing by perturbers will also produce echo-like repetitions of the null cone GW signal.
However, strong lensing occurs only when $n_E\leq 1$, while glints should be detectable in association with the null cone signal out to $n_E\gg1$.
For a fixed population of perturbers,  observable glints will be far  more numerous than strong-lensing events.

Transparent lenses obey the odd-image theorem of gravitational lensing, hence non-black-hole perturbers will produce odd numbers of images.  In contrast, each perturber produces one glint for an observer.  If all images are detectable, resolvable in time but not in direction, then glints will yield a primary signal and a single echo, while lensing will yield two or more echos.

There are other reasons to be optimistic about disambiguation of individual echo events.
Strong-lensing images have two types of time-delay---Shapiro time-delay $\Delta t_{\mathrm{Shapiro}} \sim r_S/c$ and geometric time delays, $\Delta t_{\mathrm{geom}}\sim b^2/c \ell$. For sources at cosmological distances and $b\lesssim x_E$,  $\Delta t \lesssim r_S/c$ for both.  Glints, on the other hand, have only geometric time-delays, with $\Delta t_{\mathrm{geom}}\sim n_E^2 r_S/c$.  Longer delays should make glints more readily detectable than, and more readily distinguishable from, lensed images.

Glint time delays are a simple function of the ratio of the glint-to-null-cone signal strength, 
$\mathcal{R}^{\mathrm{mid/null}}$.
The dimmer glint always follows the primary null cone signal.
In contrast,  strong-lensing image magnification ratios and time-orderings depend on the lens. 
Individual gravitational-echo events may have more than one explanation in terms of unknown perturbers, however glints and lensed images should be readily distinguished, at least statistically.

\noindent\paragraph*{Dark Matter and Glints}
Another potential source of glints is macroscopic dark matter candidates. This includes a wide range of  hypothetical compact objects composed largely of quarks or baryons, to which we refer collectively as macros \cite{Jacobs:2014yca} or
compact composite objects \cite{Zhitnitsky:2006x}.
It also includes objects farther outside the Standard Model such as axion stars and dark matter minihalos.
Gravitational-lensing searches show that compact perturbers with  $10^{-11}M_\odot\leq M_P \leq M_\odot$ have  $\Omega_P\lesssim0.03$ \cite{Carr:2020xqk}.
Still, this is an order of magnitude larger than $\Omega_{\mathrm{stars}}$, so for compact dark matter (cDM)
\begin{equation}
    N_{\mathrm{cDM}}(n_E )\gtrsim 1.5 \times 10^{-2} n_E^2 
    \frac{\Omega_{\mathrm{cDM}}}{0.03}    .
\end{equation}
Gravitational glints can therefore be used to discover such objects or stringently constrain their abundance.

\paragraph*{Summary}

Gravitational glints can be understood as the scattering of GWs off perturbations in the background (conformally) flat geometry, populating the interior of the null cone.
This results in detectable signals when GWs from conventional sources like coalescing binaries interact with the geometry sourced by conventional objects like stars, white dwarfs, or neutron stars near the LOS. 
Though we have studied these glints in the idealized case of a monochromatic GW source, a simplified perturber, and a specific symmetric geometry,
the generic conclusion is that these will cause glints  up to approximately the strength of the primary null cone signal, very briefly following it, and very nearly its faithful echos.
The precise limits of detection of such glints will depend on the specifics of the sources, the perturbers and their geometry, and will require detailed modeling.  
These are not however the echo claimed to have been detected in \cite{Abedi:2016hgu}, which is delayed compared to the primary signal by many oscillation periods, and repeats.

Glints are an automatic consequence of General Relativity, known sources and known perturbers.
More speculative compact perturbers---like a wide variety of  macroscopic dark matter candidates---would also be detectable if they form a significant fraction of the dark matter, and have masses not much smaller than $M_\odot$.

The expected number of glints per GW event is  proportional to the ratio of the null cone signal to the minimum detectable glint signal. 
It is high enough that it may be possible to extract evidence of glints from archived data, or from future data at current sensitivities.  
Future improvements to existing GW observatories are likely to result in frequently  detectable glints.

Glints will also contribute to the  stochastic GW background. 
A naive assessment would extrapolate (\ref{eqn:StoN}) and (\ref{eq:NofnE}), however we have no reason to believe that (\ref{eqn:StoN}) applies at the very large $n_E$ one would wish to include.
A compelling assessment would require us to go to at least NLO in the calculation of the signal. 

GWs thus become more than a new window on their sources. 
These sources function as the gravitational analogue of distant radar beacons, and the glints of distant perturbers like radar echoes.
Using GRAvitational  Detection  And  Ranging  (GRADAR), we can find and characterize all massive compact objects---from stars and their terminal remnants,  to hypothesized dark matter candidates and structures, and possibly planets.

Stars may fade, and dark matter may never glow, but they cannot hide from gravity.

\acknowledgements
The authors thank Klaountia Pasmatiou for her extensive contributions to the development of the ideas and calculations performed in this work. It is the authors' strongly held opinion that her refusal to be a co-author on this work is unwarranted. GDS is partly supported by a Department of Energy grant DE-SC0009946 to the particle astrophysics theory group at CWRU.
Code will be made available on request.

\bibliography{2021_PRL}

\end{document}